\documentclass{INTERSPEECH2023}
\usepackage{mathrsfs}
\usepackage{amssymb}
\usepackage{multirow}
\usepackage[table]{xcolor}
\usepackage[flushleft]{threeparttable}

\usepackage[hang,flushmargin]{footmisc} 
\usepackage{enumitem}

\interspeechcameraready

\title{Improving Label Assignments Learning by Dynamic Sample Dropout Combined with Layer-wise Optimization in Speech Separation}
\name{Chenyang Gao$^1$, Yue Gu$^2$, Ivan Marsic$^1$}
\address{
  $^1$Rutgers University\\
  $^2$Amazon APT}
\email{cg694@rutgers.edu, yguam@amazon.com, marsic@rutgers.edu}
\begin{document}
\setlength{\abovedisplayskip}{5pt}
\setlength{\belowdisplayskip}{5pt}
\setlength{\abovedisplayshortskip}{5pt}
\setlength{\belowdisplayshortskip}{5pt}
\maketitle
\begin{abstract}
In supervised speech separation, permutation invariant training (PIT) is widely used to handle label ambiguity by selecting the best permutation to update the model. Despite its success, previous studies showed that PIT is plagued by excessive label assignment switching in adjacent epochs, impeding the model to learn better label assignments. To address this issue, we propose a novel training strategy, dynamic sample dropout (DSD), which considers previous best label assignments and evaluation metrics to exclude the samples that may negatively impact the learned label assignments during training. Additionally, we include layer-wise optimization (LO) to improve the performance by solving layer-decoupling. Our experiments showed that combining DSD and LO outperforms the baseline and solves excessive label assignment switching and layer-decoupling issues. The proposed DSD and LO approach is easy to implement, requires no extra training sets or steps, and shows generality to various speech separation tasks.
\end{abstract}
\vspace{-0.5em}
\section{Introduction}
\vspace{-0.5em}
Speech separation is a specific type of source separation that focuses on separating human speech sources from overlapping speech signals. Deep learning has demonstrated great success in speech separation \cite{huang2014deep,wang2018supervised}. During the training of speaker-independent speech separation models, a commonly recognized challenge known as label ambiguity or permutation problem arises, which pertains to the ambiguity or uncertainty in assigning labels to predictions. This problem is caused by the same nature of the overlapped sound sources where the order of the labels and predictions do not match well during training. To address this issue, permutation invariant training (PIT) was introduced \cite{yu2017permutation,kolbaek2017multitalker}, which involves exploring all possible label-assignment pairs and selecting the best pair to update the model. PIT has now become the standard training approach for time-frequency (T-F) domain \cite{luo2018tasnet,luo2019conv} and time-domain \cite{luo2018tasnet,luo2019conv,luo2020dual,chen20l_interspeech,subakan2021attention,zhang2021transmask,zeghidour2021wavesplit,tzinis2020sudo,lam2021sandglasset,li2022skim} speech separation models. In time-domain approaches, 1-D convolution and transposed convolution are used as trainable front-end to replace the roles of STFT and iSTFT in T-F domain approaches, respectively.

Although PIT has effectively tackled label ambiguity for training speech separation models, it suffers from unstable label assignment switching \cite{yang2020interrupted,huang21h_interspeech,yousefi19_interspeech,tachibana2021towards}. 
This issue arises when a large proportion of label assignments abruptly switch the order in adjacent epochs, resulting in erratic training.
The unstable label assignment switching problem is claimed to be caused by the slight difference in pairwise loss during the initial epochs as in \cite{yang2020interrupted,huang21h_interspeech,yousefi19_interspeech,tachibana2021towards,YOUSEFI2023}.
Prior research attempted to address this problem from two main perspectives. Some studies \cite{yang2020interrupted,huang21h_interspeech} have combined different training strategies with the original PIT approach, including fixing label assignments \cite{yang2020interrupted} and fine-tuning a model pre-trained with speech enhancement \cite{huang21h_interspeech}, while others \cite{yousefi19_interspeech,tachibana2021towards, YOUSEFI2023} have proposed probabilistic relaxation PIT to prevent the model from being over-confident in the label assignments. However, our experiments show that these approaches are insufficient in resolving this problem, which impedes the model's ability to learn better label assignments.
To solve the above issue, we introduce a practical training approach named dynamic sample dropout (DSD) by addressing the issue of excessive label assignment switching. DSD employs a mechanism that considers previous best label assignments and evaluation metrics to identify and exclude the challenging samples that may have a detrimental effect on learning label assignments. Unlike other techniques \cite{yang2020interrupted,huang21h_interspeech}, DSD doesn't require additional data or training steps, so it's applicable in various speech separation settings. Additionally, we combine DSD with layer-wise optimization (LO) \cite{10096897, kim2022bloom, nachmani2020voice} to further enhance the model's performance. Through an extensive study of layer-wise optimization, we found that LO reduces layer-decoupling (see Section \ref{lo_results}), leading to the observed improvement in model performance. Our experiments using LibriMix data demonstrate that the proposed approach outperforms the baseline models by a significant margin in a range of 1.07 to 1.62 dB in SI-SDRi. Our contributions are:
\begin{enumerate}
\item  We assess and illustrate the limitations of current approaches in addressing excessive label assignment switching.
\item We propose a novel dynamic sample dropout strategy that employs layer-wise optimization to effectively resolve the issues of excessive label-switching and layer-decoupling without the need for additional training samples or steps. 
\item We carry out extensive experiments to demonstrate the consistent performance enhancement and generality of the proposed dynamic sample dropout with the layer-wise optimization approach across various speech separation tasks.
\end{enumerate}
\section{Related Work}
\vspace{-0.5em}
\subsection{Label ambiguity and Permutation invariant training}
\label{permutation_invariant_training}
\vspace{-0.5em}
The problem of mono-channel speech separation is formulated as follows. Given a mixture speech signal $X$ containing $N$ speakers: $X= \sum_1^Ns_i+n,s_i\in R^t$, where $s_i$ is the clean source of speaker $i$ and $n$ is background noise. The goal is to recover the speech for each speaker from the mixture waveform. Considering a two-speakers case, the label ambiguity problem occurs because the model’s predictions $r_1$ and $r_2$, and labels $s_1$ and $s_2$, could be matched arbitrarily. That is, $r_1$ could be the estimated recovery of either $s_1$ or $s_2$. Consequently, there exists $N!$ possible different combinations of prediction-label pairs; the choice in different combinations is known as the problem of label ambiguity. Permutation invariant training (PIT) \cite{yu2017permutation, kolbaek2017multitalker} has been proposed to solve this label ambiguity during training. In short, PIT traverses all the possible prediction-label pairs and selects only the optimal one to update the model. Although PIT achieves great success in training speech separation models, the unstable label assignment switching during training impairs the model’s performance.
\vspace{-0.5em}
\subsection{Learning better label assignments}
\vspace{-0.5em}
Different approaches have been proposed to overcome unstable label assignment switching in the following two ways.

\textbf{Improving training strategy}\textemdash The first group tries to improve the training strategy of speech separation with the original PIT. In \cite{yang2020interrupted}, the authors proposed a strategy called interrupted and cascaded training. It avoids label assignment switching by fixing the label assignments after the initial PIT training epochs. 
In \cite{huang21h_interspeech}, the authors proposed using speech enhancement as the pre-training task to stabilize the label assignments. 

\textbf{Improving PIT}\textemdash The other group tried to modify the original PIT. ProbPIT \cite{yousefi19_interspeech}, SinkPIT \cite{tachibana2021towards}, and soft-minimum PIT \cite{YOUSEFI2023} were proposed as the probabilistic relaxation version of the original PIT. They used a weighted sum of losses over all possible permutations to avoid the model being over-confident to a specific permutation, showing better learning label assignments ability than the original PIT. 

We performed the ablation study to assess if the excessive label assignment switching problem is well-addressed with interrupted and cascaded training strategy \cite{yang2020interrupted} and SinkPIT \cite{tachibana2021towards}. 
However, we still observed a sudden excessive switching during training with these approaches, indicating that excessive label assignment switching is not well addressed and still hinders the model from learning better label assignments (see Section \ref{baseline_exp}).
\begin{figure}[b]
  \vspace{-1.5em}
  \centering
  \includegraphics[width=\linewidth]{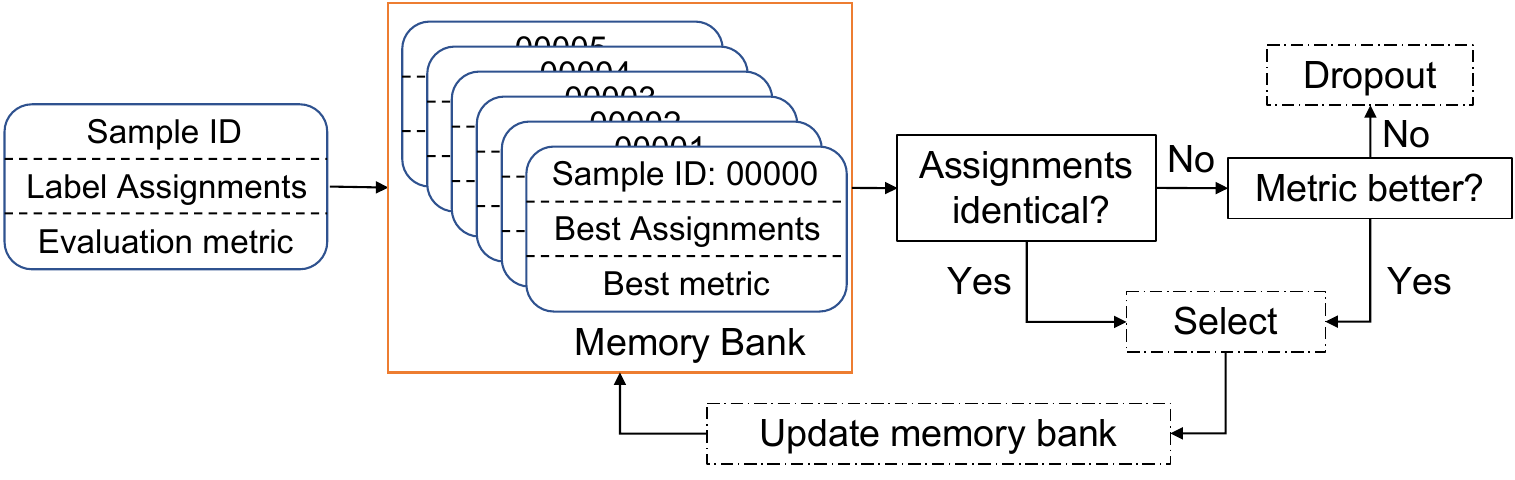}
  \caption{The proposed dynamic sample dropout.}
  \label{fig:dsd}
   \vspace{-1.5em}
\end{figure}
\vspace{-1.0em}
\section{Dynamic Sample Dropout and Layer-wise Optimization}
\subsection{Dynamic sample dropout}
\vspace{-0.5em}
We introduced our dynamic sample dropout (DSD) training method to overcome the problem of excessive label assignment switching (see Figure \ref{fig:dsd}). During our reproduction of previous methods and baselines, we noticed that the evaluation metrics would abruptly decrease after certain training samples. Our hypothesis for this phenomenon is that challenging training samples can negatively impact the learned label assignments and result in excessive label assignment switching and inconsistent training progress. To validate and resolve this issue, we introduce the DSD training strategy, which dynamically removes filtered training samples based on an evaluation of both the metric and past label assignments. This approach uses a memory bank to keep track of the best evaluation metric and corresponding label assignments for each training sample. The challenging samples are omitted during the corresponding training iterations to maintain stable label assignments. The memory bank is initialized at the first training epoch, where it records the label assignments and evaluation metrics for every sample. In the remaining training epochs, the DSD uses the following criteria to select or dropout the samples at each optimization step:
\begin{enumerate}
\item  Select: \romannumeral1) if the current label assignment chosen by PIT is identical to the recorded best assignments; \romannumeral2) if the label assignment order changed, the evaluation metric relaxed outperforms the recorded best evaluation metric. We update the record for these training samples in the memory bank.
\item Dropout: if the label assignment order changes, the evaluation metric is not relaxed better than the recorded best evaluation metric. The loss for this training sample will drop out.
\end{enumerate} 
The criterion for ``relaxed better'' is defined as:
\vspace{-1.0mm}
\begin{align}
  M_{cur}*(1+ sgn(M_{cur})*\epsilon)>M_{best}
\end{align}
where $M_{cur}$ represents the current evaluation metric and $M_{best}$ represents the best evaluation metric, $\epsilon$ is a relaxation factor.
The $sgn$ function ensures fair sign comparison for negative evaluation metrics.
The relaxation step enables DSD to tolerate samples that result in a slightly worse evaluation metric but switch the label assignments. DSD discards challenging samples that may disrupt the learned label assignments during the training process, thereby maintaining a stable label assignment switching ratio. Instead of discarding, an alternative approach is to persist with the previously recorded best label assignments for these challenging samples. Specifically, we use the best-recorded label assignments stored in the memory bank to recalculate the loss for these challenging samples. The \textit{reorder} operation insists on the best-recorded label assignments for these challenging samples and allows them to still participate in the training process. We refer to this approach as DSD (\textit{reorder}) to distinguish it from DSD (\textit{dropout}).
\vspace{-1.0em}
\subsection{Layer-wise optimization}
\vspace{-0.5em}
We further combined the proposed dynamic sample dropout (DSD) with layer-wise optimization (LO) to enhance the learning of label assignments. LO was introduced for efficient inference in previous studies, such as in \cite{10096897, kim2022bloom, nachmani2020voice}, where intermediate layers are trained directly with the target, allowing for early-exit strategies to save inference time. Specifically in speech separation, for a model with $N$ repeated sequential modeling blocks, the intermediate outputs from each layer have the same shape, and they are used to reconstruct the clean target. Layer-wise optimization computes a loss term for each layer and sums them up. The layer-wise optimization for speech separation is shown as follows:
\vspace{-1.0mm}
\begin{align}
  \mathscr{L}=\frac{1}{N}\sum_{i=1}^Nw_i*\text{PIT}(\Tilde{S}_i, S)
\end{align}
where $\Tilde{S}_i$ is the reconstructed source from intermediate layer $i$, $S$ is the target source, and $w_i$ is a weighted scalar that controls each loss's contribution term to the final loss. In addition to the inference efficiency, the layer-wise optimized model also outperforms the model that is only optimized by a single loss term from the last layer, as shown in \cite{nachmani2020voice}. Instead of giving a subjective explanation for the improvement, we conducted experiments and analyses to demonstrate how the improved gradient flow affects the behavior of each intermediate layer, by comparing the dissimilarity between the label assignment switching ratio curve of intermediate layers and the last layer. We used the same approach as in \cite{nachmani2020voice}, where it shares the weights for the mask estimation network and decoder for intermediate outputs generation. We believe this discrepancy is the actual reason for the performance improvement by using LO. Our aim is to increase the popularity of layer-wise optimization in the speech separation community with this new insight.
\begin{table}[!b]
  \vspace{-1.0em}
  \caption{Performance on the \textit{sep\_clean} test set. ``L'' indicates the number of epochs trained with PIT.}
  \vspace{-0.5em}
  \label{tab:baseline_res}
  \centering
  \begin{tabular}{cc}
    \hline\hline
    \textbf{Method} & \textbf{SI SDRi/SDRi} \\
    \hline
    PIT= DSD ($\epsilon=+\infty$) &	$15.13/15.53$ ~~~ \\
    \rowcolor{gray!20} SinkPIT	                      & $15.28/15.70$ ~~~ \\
    (PIT)-(fix) \textit{L=1}      &	$14.98/15.39$ ~~~ \\
    (PIT)-(fix) \textit{L=5}	  & $14.59/15.00$ ~~~ \\
    (PIT)-(fix) \textit{L=10}	  & $14.65/15.05$ ~~~ \\
    (PIT)-(fix) \textit{L=25}	  & $14.99/15.40$ ~~~ \\
    (PIT)-(fix) \textit{L=50}	  & $14.84/15.24$ ~~~ \\
    (PIT)-(fix) \textit{L=75}	  & $14.93/15.34$ ~~~ \\
    (PIT)-(fix) \textit{L=100}	  & $14.79/15.19$ ~~~ \\
    (PIT)-(fix) \textit{L=150}	  & $14.92/15.31$ ~~~ \\
    \hline
    DSD (\textit{reorder}, $\epsilon=0.0$)	          & $15.46/15.90$ ~~~\\
    DSD (\textit{dropout},$\epsilon=0.0$)	      & $15.62/16.07$ ~~~\\
    \rowcolor{gray!20} DSD (\textit{dropout},$\epsilon=0.1$)	      & $15.75/16.20$  ~~~\\
    DSD (\textit{dropout},$\epsilon=0.2$)	      & $15.59/16.05$~~~ \\
    DSD (\textit{dropout},$\epsilon=0.5$)	      & $15.39/15.83$ ~~~ \\
    \hline
    \rowcolor{gray!20} LO	      & $15.84/16.27$ \\
    \hline\hline
  \end{tabular}
  \vspace{-1.0em}
\end{table}

\section{Experimental Setup and Implementation}
We used DPTNet \cite{chen20l_interspeech} as our baseline model for its competitive performance in various speech separation and enhancement tasks \cite{chen20l_interspeech, dang2022dpt,wang2021tstnn}. The DPTNet is a time-domain masking-based model that uses dual-path processing \cite{luo2020dual}, which segments audio into short segments and then sequentially applies the intra and inter-procedure. The dual-path processing allows the model to handle the long-range dependencies in long audios. We used the default configuration as in \cite{chen20l_interspeech} to construct the separation model, where the number of the improved transformer is set as six. We used a kernel size of 16 and strided with 8 in encoder/decoder as in \cite{luo2019conv,luo2020dual}. The entire model contains 2.7 M trainable parameters.

We employed the open-source LibriMix dataset \cite{cosentino2020librimix}, derived from the Librispeech dataset \cite{panayotov2015librispeech}, for our experiments. We conducted experiments with various subsets, including different numbers of speakers (Libri2Mix and Libri3Mix) and varying conditions (clean or noisy). The results were reported on the minimum version of the corresponding test sets, following previous studies \cite{cosentino2020librimix}.

We built the model using the Asteroid toolkit \cite{pariente20_interspeech}. To ensure a fair comparison, the model was trained for 200 epochs with a batch size of 24 in all different experiment settings, consistent with previous work \cite{huang21h_interspeech}. The Adam optimization algorithm \cite{DBLP:journals/corr/KingmaB14} was used with an initial learning rate of 1e-3, and the gradients were clipped with a maximum $L_2$ norm of 5. The patience for halving the learning rate was set to 10 for the first 80 epochs and 5 for the remaining epochs. The audio segments were divided into 3-second segments for both training and validation. The scale-invariant signal-to-distortion (SI-SDR) metric was used as the training objective \cite{le2019sdr} and the evaluation metric in DSD. We set the $w_i=\frac{\text{layer index}}{\text{total blocks}}$ in LO. We evaluated the improved signal quality using the SI-SDRi and SDRi metrics.
\vspace{-2.0mm}
\begin{figure}[!t]
  \vspace{-1.0em}
  \centering  \includegraphics[width=\linewidth]{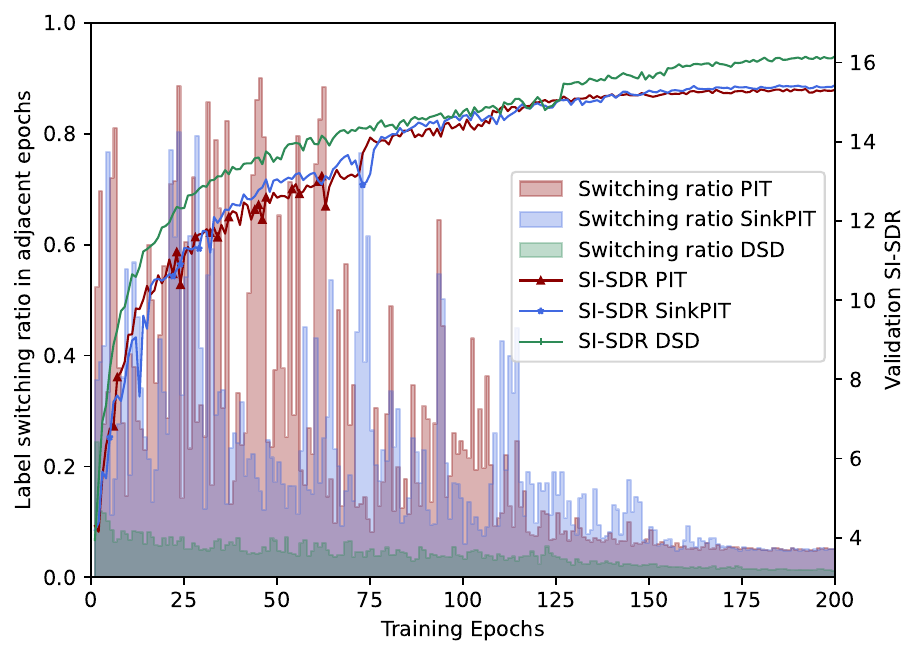}
  \caption{Label assignments switching ratio versus validation SI-SDR for PIT, SinkPIT, and DSD.}
  \label{fig:lsw_basaeline}
  \vspace{-2.0em}
\end{figure}
\section{Experiment Results and Analysis} 
\subsection{Baseline experiments with PIT and its variants} \label{baseline_exp}
\vspace{-1.0mm}
We conducted experiments in three settings: \romannumeral1) PIT; \romannumeral2) SinkPIT; and \romannumeral3) Interrupted \& Cascaded (only PIT-(fix) step), using the LibriMix train-100 subset with the \textit{sep\_clean} task. SinkPIT demonstrates the best performance (Table \ref{tab:baseline_res}), achieving 15.28 dB in SI-SDRi compared to the original PIT (15.13 dB). In contrast to the results in \cite{yang2020interrupted}, PIT-(fix) shows the worst results with a difference ranging from 14.59 to 14.98 dB in SI-SDRi, depending on the value of \textit{L}.

To better understand the training process, we analyzed the label assignment switching ratio curve as follows. We selected the label assignment using PIT for each training sample after every epoch and defined the fraction of samples with different label assignments in adjacent epochs as the label assignment switching ratio. Figure \ref{fig:lsw_basaeline} shows the label assignment switching curve against validation SI-SDR for PIT and SinkPIT. We noted that a large fraction of label assignments switched in adjacent epochs,
causing instability in the training path and a decline in SI-SDR, as indicated by the markers on the SI-SDR curves. 
This observation also sheds light on the reason behind the ineffectiveness of PIT-(fix); excessive label assignment switching in the initial PIT procedure cannot guarantee that the model learns promising label assignments, thus fixing the label assignments in later steps cannot enhance the performance (\textit{L}=1 outperformed \textit{L}=100). 
These findings suggest that excessive label assignment switching hinders the model's ability to learn better label assignments, and existing approaches like Interrupted \& Cascaded and SinkPIT do not effectively solve this issue.
\vspace{-1.5mm}
\subsection{Dynamic sample dropout}
\vspace{-1.0mm}
We evaluated the proposed dynamic sample dropout (DSD) approach using the LibriMix train-100 subset with the \textit{sep\_clean} task. It is important that when $\epsilon=+\infty$, the DSD strategy becomes identical to PIT because it will accept all of the data. As shown in Table \ref{tab:baseline_res}, we observe a significant improvement in performance compared to the baseline when we used DSD (15.75 versus 15.13 in SI-SDRi). We then performed an ablation study comparing DSD (\textit{reorder}) and the original DSD (\textit{dropout}) and found that DSD (\textit{dropout}) outperformed DSD (\textit{reorder}), indicating that \textit{dropout} is a more efficient method in dealing with challenging samples. We conjectured that DSD (\textit{reorder}) still involved challenge samples in training, which impeded learning better label assignments. Moreover, the speech separation performance initially improved and then decreased as the relaxation factor was increased, suggesting that relaxation is more effective in handling the criteria for dropping training samples. We also evaluated the label assignment switching curve of the proposed DSD method (as shown in Figure \ref{fig:lsw_basaeline}). The label assignment switching ratio of the model trained with DSD became significantly more stable compared to the original PIT and SinkPIT. The faster convergence rate and improved performance further highlighted the importance of maintaining a stable label assignment switching ratio for the model to learn better label assignments. Additionally, in the most stringent scenario ($\epsilon=0.0$), the percentage of discarded samples in the training set gradually decreased from $3\%$ to less than $1\%$ over the course of training, pointing to the fact that challenging samples disrupt the learned label assignments during training and result in excessive label assignment switching. For subsequent experiments, we employed $\epsilon=0.1$ and \textit{dropout} for DSD.

\vspace{-1.0mm}
\begin{table}[!tb]
\vspace{-1.0mm}
\begin{threeparttable}
\caption{Similarity between intermediate layers and the last layer. $L_1$ distance is reported.}
\vspace{-1.0mm}
  \label{tab:simiarity_comparison}
  \centering
\begin{tabular}{lccc|>{\columncolor[gray]{0.9}}c}
  \hline\hline
  \textbf{Comparison} & \textbf{PIT} & \textbf{DSD} & \textbf{LO} & \textbf{DSD+LO}\\
\hline\hline
  1 vs 6	& 37.50	& 33.31 & 27.50 & 0.83 ~~~ \\
  2 vs 6	& 29.98	& 16.73 & 2.26 & 0.20 ~~~ \\
  3 vs 6	& 25.56	& 4.16 & 0.13 & 0.07 ~~~ \\
  4 vs 6	& 14.78	& 0.56 & 0.05 & 0.03 ~~~ \\
  5 vs 6	& 6.67	& 0.37 & 0.03 & 0.02 ~~~ \\
\hline\hline
\end{tabular}
\end{threeparttable}

\vspace{-1.5em}

\end{table}

\vspace{-1.0mm}
\begin{figure}[b]
  \vspace{-1.5em}
  \centering
  \includegraphics[width=\linewidth]{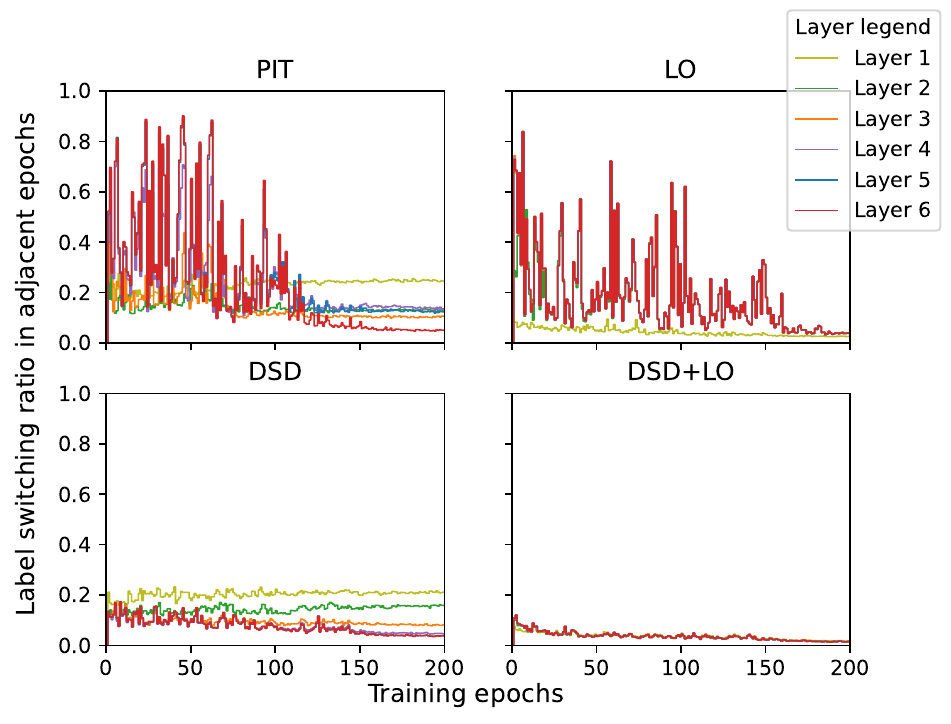}
  \caption{Layer-wise label assignment switching curve with different training strategies.}
  \label{fig:layer-wise switching ratio}
   \vspace{-1.5em}
\end{figure}

\subsection{Layer-wise optimization}
\label{lo_results}
\vspace{-0.5em}
We evaluated the performance of the layer-wise optimization (LO) approach on the LibriMix train-100 subset with the \textit{sep\_clean task}. The results show that using LO improved the performance (Table \ref{tab:baseline_res}), achieving 15.84 SI-SDRi and 16.27 SDRi, surpassing the baseline results of PIT and SinkPIT. We analyzed the label assignment switching ratio curve for each layer of the models trained with and without LO. We determined how the improved gradient flow in LO affects the behavior of the intermediate layers by comparing the similarity in the label assignment switching curves of the intermediate layers with that of the last layer. We argue that this similarity somehow reflects the training process of the model. We also indicate this similarity by calculating the $L_1$ distance between label assignment switching curves. Based on Figure \ref{fig:layer-wise switching ratio} and Table \ref{tab:simiarity_comparison}, we made the following observations:
\begin{enumerate} 
\item The curves of the label assignment switching ratio for Layers 2-6 change in tandem in the LO-trained model, while Layer 1 shows a distinct trend (Figure \ref{fig:layer-wise switching ratio}, LO.) 
\item Unlike LO, where Layers 2-3 showed a similar trend in the changing of label assignment switching ratio curves, only Layers 4-6 are changing in unison in PIT and DSD, even though they were trained without LO (Figure \ref{fig:layer-wise switching ratio}, PIT \& DSD.) 
\item The problem of excessive label assignment switching exacerbates the differences in the label assignment switching ratio curves, with the distance of PIT-trained model being much greater than DSD-trained model (Table \ref{tab:simiarity_comparison}, PIT \& DSD.) 
\end{enumerate}
The resemblance and disparity between the label assignment switching ratio curves indicate coherence and incoherence, respectively, in the training directions of each intermediate layer during training. We refer to the issue of intermediate layers having dissimilar switching ratio curves as the ``layer-decoupling'' problem. Our findings suggest that LO serves as a regularization technique for the training direction of the middle layer, which significantly mitigates the layer-decoupling problem. Nevertheless, the excessive problem of label assignment switching still affects the layer-decoupling problem in the LO-trained model (Layer 1 has a different trend), and further amplifies it (observation \#3). 
\begin{table}[!tb]
\resizebox{\columnwidth}{!}{
\begin{threeparttable}
\caption{Experiment results on different tasks, results reported on corresponding min version of test sets. For DSD, we set $\epsilon=0.1$}
\label{table:additional_results}
\centering
\begin{tabular}{lcccccccc}
  \hline\hline
 \multirow{1}{*}{\textbf{Task}} && \multirow{1}{*}{\textbf{DSD}} && \multirow{1}{*}{\textbf{LO}} && \multirow{1}{*}{\textbf{train-100}} && \multirow{1}{*}{\textbf{train-360}}\\
 \hline\hline
 2spk-C (PIT) && $\times$ && $\times$ && $15.13/15.50$ && $15.92/16.27$~~~\\ 
 2spk-C && $\checkmark$ && $\times$ && $15.75/16.20$ && $16.70/17.09$~~~\\ 
 2spk-C && $\times$ && $\checkmark$ && $15.84/16.27$ && $16.84/17.19$~~~\\ 
 \rowcolor{gray!20} 2spk-C (Ours) && $\checkmark$ && $\checkmark$ && $16.28/16.75$ && $17.22/17.56$~~~\\ 
 \hline
 2spk-N (PIT) && $\times$ && $\times$ && $11.64/12.21$ && -/-~~~\\ 
 2spk-N && $\checkmark$ && $\times$ && $12.33/12.92$ && -/-~~~\\
 2spk-N && $\times$ && $\checkmark$ && $12.50/13.10$ && -/-~~~\\ 
 \rowcolor{gray!20}2spk-N (Ours)&& $\checkmark$ && $\checkmark$ && $12.79/13.39$ && -/-~~~\\ 
 \hline
 3spk-C (PIT) && $\times$ && $\times$ && $11.92/12.38$ && $13.34/13.76$~~~\\
 3spk-C && $\checkmark$ && $\times$ && $12.54/13.00$ && $14.10/14.53$~~~\\ 
 3spk-C && $\times$ && $\checkmark$ && $12.58/13.04$ && $14.20/14.64$~~~\\ 
 \rowcolor{gray!20}3spk-C (Ours)&& $\checkmark$ && $\checkmark$ && $12.99/13.47$ && $14.96/15.40$~~~\\ 
 \hline\hline
\end{tabular}
\end{threeparttable}
}
\vspace{-1.5em}
\end{table}
\vspace{-2.0mm}
\subsection{Combining DSD and LO}
\vspace{-1.0mm}
To take advantage of both the DSD and LO strategies, we combined them to address the excessive label assignment switching and layer-decoupling problems simultaneously. We first applied the same experiment and the analysis as in Section \ref{lo_results}. Figure \ref{fig:layer-wise switching ratio} and Table \ref{tab:simiarity_comparison} shows that the DSD+LO eliminates both excessive label assignment switching and layer-decoupling problem. And it also leads to a further improvement in the separation performance (in Tale \ref{table:additional_results}). To show the applicability of DSD+LO in general speech separation, we evaluated the performance of the combined DSD+LO approach on various speech separation tasks, including a larger dataset (train-360 subset), noisy conditions (\textit{sep\_noisy}), and more speaker scenarios (Lirbi3Mix). The results (Table \ref{table:additional_results}) show that the proposed DSD+LO-trained models outperform the baselines with a margin between 1.07 to 1.62 dB in the SI-SDRi. The DSD+LO-trained models also outperform the DSD- and LO-trained models, showing the complementarity of these two training strategies. 
\vspace{-1.0em}
\section{Conclusion}
We studied the issue of excessive label assignment switching in speech separation and discovered that existing methods were unable to effectively address it. We proposed dynamic sample dropout (DSD) to maintain a stable label assignment switching ratio by removing samples that could negatively impact the learned label assignments. We further introduced layer-wise optimization (LO) to improve separation performance by reducing layer decoupling. By combining DSD and LO, our proposed model outperformed all baselines, effectively addressing both excessive label assignment switching and layer decoupling. The DSD + LO training strategy is easy to implement, requires no extra training sets or steps, and demonstrates strong generality to various single-channel speech separation tasks. We believe that it could be easily employed in multi-channel scenarios because PIT is also widely used in multi-channel settings.
\noindent
\let\thefootnote\relax\footnote{This work is supported by NIH/NLM grant number R01LM011834 and NSF grant IIS-1763827.}
\bibliographystyle{IEEEtran}
\bibliography{mybib}

\end{document}